\documentclass[conference]{IEEEtran}
\IEEEoverridecommandlockouts
\usepackage{cite}
\usepackage{amsmath,amssymb,amsfonts}
\usepackage{algorithmic}
\usepackage{graphicx}
\usepackage{textcomp}
\usepackage{gensymb}
\usepackage{xcolor}
\def\BibTeX{{\rm B\kern-.05em{\sc i\kern-.025em b}\kern-.08em
    T\kern-.1667em\lower.7ex\hbox{E}\kern-.125emX}}
\begin{document}

\title{An Embedded System for Monitoring Industrial Air Dehumidifiers using a Mobile Android Application for IEEE 802.11 Networks\\
}

\author{\IEEEauthorblockN{Erik de Oliveira Rosa, Lincoln Grabarski, \\ Marcos F. Fragoso, Allan C. Krainski Ferrari}
\IEEEauthorblockA{\textit{University Center of Arauc\'{a}ria (UNIFACEAR)$ ^1 $}\\
Arauc\'{a}ria, Paran\'{a}, Brazil \\
\{edioliver2011,marcos\_fernando120\}@hotmail.com, \\ \{lincolngrabarski,allanckferrari87\}@gmail.com}
\and
\IEEEauthorblockN{Jefferson Rodrigo Schuertz and Carlos \\ Alexandre Gouvea da Silva$ ^1 $}
\IEEEauthorblockA{\textit{Electrical Engineering Department (DELT)} \\
\textit{Federal University of Paran\'{a} (UFPR)}\\
Curitiba, Paran\'{a}, Brazil \\
jeffersonschuertz.eng@gmail.com, carlos.gouvea@ieee.org}
}


\maketitle

\markboth{XXXVIII SIMPÓSIO BRASILEIRO DE TELECOMUNICAÇÕES E PROCESSAMENTO DE SINAIS - SBrT 2020, 13--16 DE SETEMBRO DE 2020, FLORIANÓPOLIS, SC}{}

\begin{abstract}
The constant technological evolution allowed signi-ficant advances and improvements in the processes of industries, mainly in areas that demand greater control and environmental air efficiency. In this way, Embedded Systems allow the deve-lopment of products and services that aim to solve or propose solutions in these industrial environments. This article presents the development of an Embedded System with a Programmable Logic Controller (PLC) and Arduino for industrial air dehumi-difier, which allows the monitoring of failures remotely from a reliable data communication in a mobile application for Android operating system (OS) on a wireless network IEEE 802.11. As result, a prototype of the test bench for the Embedded System is presented in which the main parameters of temperature sensors and operating conditions of the dehumidifiers are checked.

\end{abstract}


\section{Introduction}
Air quality is one of the decisive factors in the control of environments that require specific care, e.g. in hospitals or outpatient clinics for the isolation of microorganisms and mitigation of the occurrence of infections.
Several pathogens found in air conditioning units highlighting the need for air quality control measures in air-conditioned hospital environments \cite{Afonso:2004}. 
However, air control isn't limited to biological environments, but also to residential and industrial environments using specific equipment, such as dehumidifiers.

Industrial automation of air quality has evolved and provided an increasing number of techniques to control temperature, humidity and removal of polluting waste industrial environments.
There are many consequences of poor air quality and can cause respiratory diseases in people. 
In an industrial environment this quality can generate a greater number of absences due to illness or reduction in work capacity of employees.
The dehumidifiers are generally important to maintain proper humidity in the environment, especially in those where such control is required by technical standards and external regulatory, or internal procedures of industries.
Usually, the air dehumidifier is used to generate air at low relative humidity (RH) and ensure quickly and suitably the cooling process \cite{Faten:2016}.

In industrial environments, dehumidifiers represent an ally in temperature and RH control. 
In many industrial processes, the control of these air conditions can affect the quality of the products produced and also in the way they are stored. In food industries, these conditions are important to dry some types of food \cite{Sosle:2003}\cite{Moses:2014}, supermarket applications \cite{Capozzoli:2006}, thermal comfort of people \cite{Mazzei:2005}, among other.
For humidity sensitive applications such as pharmaceutical, and chemical industries it is required very dry air dehumidifiers because is difficult to achieve and energy inefficient using conventional systems \cite{Rafique:2016}.
However, these machines also are used in agricultural process \cite{Yang:2013}, greenhouse projects \cite{Zapata:2019}, among other.
At industrial sector, the main type of dehumidifiers are based at desiccant materials in which are used for absorption of water (humidity) of air.

The objective of this work is to implement an embedded monitoring system for industrial air dehumidifiers that helps users to reduce downtime in their process in cases of corrective maintenance. 
This embedded system assists in the control and verification of the need for preventive maintenance, so that can plan a programmed shutdown of the dehumidifier.
The monitoring is done through a mobile application developed for smartphones with Android OS and to be used in wireless local area networks (WLAN).

\section{Related Works}

Even existing since the 1980s' the air dehumidifiers are receiving varied updates and applications through simple or sophisticated embedded systems. These embedded systems make the monitoring and process control more effective and efficient in which: it is possible to check the status of dehumidifiers remotely through mobile devices, send notifications of failures in these equipments, improve the performance and efficiency of dehumidifiers and control the levels of air quality through remote systems.

In \cite{Lee:2015} the authors suggests an industrial dehumidifier automatic controlling system that automatically maintains optimized humidity using smart phone connected under a wireless network, in which it provides optimized temperature and humidity environment in growing environment of plants and animals. 
The system is based a small-scale sensor network with temperature and humidity sensor installed in air dehumidifier.
A mobile application allow turn on or turn off the equipment and also verify the temperature in real-time, humidity of air, luminosity inside the environment where the equipment is working, and an information indicative of working of dehumidifier.

A residential application developed by \cite{Morimoto:2013} implemented a coordination of two dehumidifiers based on humidity sensor information in the distributed manner, but other devices also are controlled using a smart outlet network. 
A novel remote-controlled system for use with air-conditioning units based in solid desiccant dehumidifiers which can be employed to create low carbon emission buildings is proposed by \cite{Peng:2019}. 
The system employs multiple information technologies such as sensor fusion, digital input/output communication and mobile technologies to monitor and control the internal conditions of buildings.
A mobile applicative is employed by \cite{Ding:2015} provides a control system to adjust parameters of air conditioners, humidifiers, dehumidifiers, air purifiers and light via extended interfaces.
An Internet of Things (IoT) application proposed by \cite{Prakasa:2019} monitors regularly six Arduino sensor in order to trigger dehumidifier, humidifier, and air purifier to maintain air quality using a mobile device.

Although the previous works present a set of solutions for controlling and monitoring air dehumidifiers through mobile applications, it still lacks work for embedded systems for industrial environments.

\section{Proposed System}

The proposed Embedded System to monitor the air quality and work conditions of dehumidifiers is based at a set of machines from the Munters Company.
Munters is a global leader in energy-efficient air treatment and climate solutions.
Munters industrial dehumidifiers provide to clients a set of essential dehumidifying functions, with additional modules or other functions, such as filtration, cooling, heating and air mixing. 
These additional modules integrate with other components of the air dehumidification process, requiring connections, such as cooling pipes, ducts and sensors.   

Fig. 1 presents the diagram of proposed system, and is described as follow.

\begin{figure}[!ht]
	\centering 
	\includegraphics[trim = 5mm 5mm 5mm 5mm, clip, width=8.9cm]{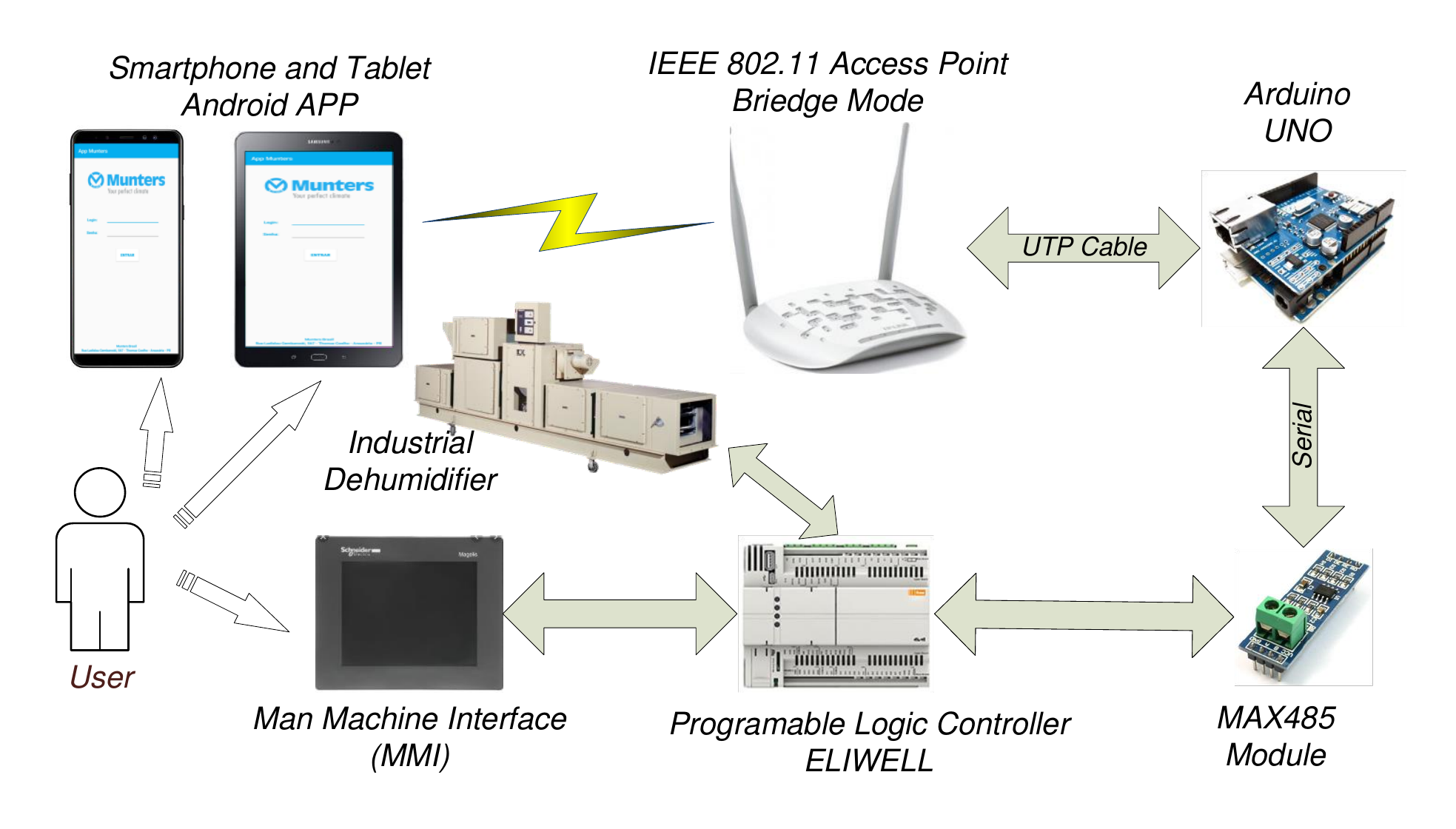} 
	\caption{Monitoring Dehumidifier System Diagram.}
	\label{figura:poligonalrota}
\end{figure}

\subsection{Programable Logic Controller (PLC)}

The main element to monitor and verify the operating conditions of the dehumidifier is the PLC ELIWELL AVC126006I500 developed by Scheneider Company.
It is connected to the dehumidifier and is responsible for reading the various operating parameters of the equipment through the 12 analogical and 12 digital input ports.
The configuration of the reading parameters of inputs and outputs is done through proprietary software from the manufacturer of the PLC.
The different connectivity modes allow the PLC to communicate with other important devices in the development of this work.
The PLC is connected on the Ethernet interface to the Man Machine Interface (MMI), via an unshilded twisted pair (UTP) cable.
Through the MMI (Fig. 2) with 8'', the local network parameters are configured, such as the PLC device IP address, the local network address, the mask address and the TCP/IP port number (502) for access via the mobile application.
In this work the interaction with MMI is just for maintenance support people and not by common users.  

\begin{figure}[!ht]
	\centering 
	\includegraphics[width=5cm]{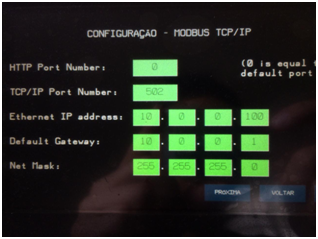} 
	\caption{Man Machine Interface (MMI).}
\end{figure}

One of the two RS485 interfaces (Modbus RTU) is used, the PLC is connected to the arduino through a MAXRS485 module.
The MAX485 module allows communication via a serial cable between the PLC and an Arduino Uno microcontroller in which it will have the embedded monitoring system.

The arduino is not used to read information of sensor because the main device to read and receive these informations is the PLC.
PLC had a set of solution in order to provide an improved controlling and management of dehumidifier.
Still, the PLC do not provide support to connect communication devices for wireless IEEE 802.11 interface in order to allow connecting mobile devices.


\subsection{Embedded System at Arduino Uno}

An application developed using C programming language was stored on the microcontroller Arduino in order to receive the information from the dehumidifier sensors through the PLC.
The information received by PLC is stored in memory elements named registers and within these registers, there are subsets that perform a function depending on their characteristic.
In this work, holding registers were used, where their main function is to read and write data with a 16-bit data message at a logical address between 40001 to 49999.

For programming it was added to the modbusslave library, in which it represents an excerpt of the previously ready modbus software. Then, all lines of code were written via the programming line in which the necessary parameter values ​​were requested. For each type of data accessed, the correct register for accessing that data is associated, as shown in Table I. 
Each register code represents the address of data register in PLC with information about dehumidifier conditions.

\begin{table}[htbp]
	\caption{Register Code List}
	\begin{center}
	\begin{tabular}{|c|c|c|}
		\hline
		\textbf{Register code} & \textbf{Register PLC} & \textbf{Read information} \\ \hline
		MB\_4000 & A0	& Test register \\ \hline
		MB\_4001 & SRT  & Reactive outlet temperature                          \\ \hline
		MB\_4002 & PA   & Reactive post-heating temperature                          \\ \hline
		MB\_4003 & ST1  & Process input temperature                          \\ \hline
		MB\_4004 & SU1  & Moisture Intake Process                          \\ \hline
		MB\_4005 & ST2  & Process output temperature                          \\ \hline	
		MB\_4006 & SU2  & Process output humidity                          \\ \hline
		MB\_4007 & PRE  & Pre-cooling temperature                          \\ \hline		
		MB\_4008 & POS  & Post-cooling temperature                          \\ \hline
		MB\_4009 & PST1 & Pressure Sensor                          \\ \hline
	\end{tabular}
	\end{center}
\end{table}

In industrial plants there are a lot of machinery and processes that naturally generate high noise that can result in high packet loss rates \cite{Silva:2018}. 
For this reason it was used for communication Arduino Ethernet module to the application, the back more reliable cable for data transmission than other types. The module is fitted over the Arduino and for this reason does not need any external connection.
An Access Point (AP) configured as bridge mode is connected to the Arduino Uno via UTP cable in order to operate only as a data transmission antenna.
The AP is configured to match the IEEE 802.11b/g/n standard.
IEEE 802.11, known as WiFi, has become the main standard for WLAN and was chosen due to ease of configuration and become is one of the most important ways to connect devices to the Internet, therein improving productivity and encouraging information sharing \cite{Silva:2019}.
Including WiFi communication, it is possible to use alternative technologies such as Bluetooth or ZigBee (IEEE 802.15.4) \cite{Silva:2017}.

\subsection{Android Applicative Mobile}

For the development of the mobile application, Android Studio was used as an integrated development environment (IDE) and a smartphone with Android 9.0 version.
The construction of the app started with the login screen, where the user and password data are inserted, as a form of reliability so that only authorized users can access the system.
In order to improve the usability of users, usual rules for creating screen designs based on software quality standards such as ISO 9126 were considered.
API sockets were used to establish a secure connection via TCP/IP protocols for communication between the mobile application and the Arduino.

\section{Results}

The results are presented in terms of the application and testing of the system in the plant created to present the embedded system in operation.
In this work it was not possible to carry out the tests in a real scenario, however the same PLC was used, which is generally used in the control of this equipment in the industrial plant.
In this case, the results in a controlled scenario is expected they will be adequate.

The Fig. 3 presents the bench developed for the Embedded System air monitoring. 
This bench consists of five main parts: the MMI (1) showing the equipment data; the set of LEDs, potentiometers and keys (2) which have the function to simulate failures and alerts the machine; the circuit breakers (3) which protect the system against overload and short circuit; PLC (4) being fed by supply 110 $ \text{V}_{\text{AC}} $ to 24 $ \text{V}_{\text{DC}} $, is responsible for all the logic of the equipment; finally the Arduino (5) contains all logic charge to receive PLC data and transmit to the smarthphone along with the MAX485 (5) that converts the RS485 signal to TTL.

\begin{figure}[!ht]
	\centering 
	\includegraphics[width=6cm]{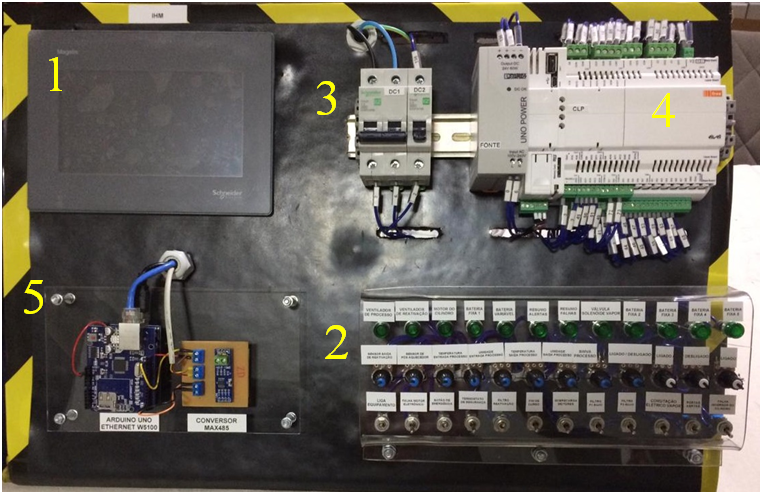} 
	\caption{Front view of Bench.}
\end{figure}

For the validation of the monitoring system of industrial air dehumidifier, data transmission tests were performed between the PLC and the Arduino through the MAX485 module. 
The data received in the Arduino using the Ethernet shield are sent packets to the Android application.
As expected, the application has received the information that was requested from the PLC and operating tests were performed a dehumidifier. These results are presented below.

\begin{figure*}[!ht]
	\centering
	\begin{tabular}{c c c}
		\includegraphics[width=3.8cm]{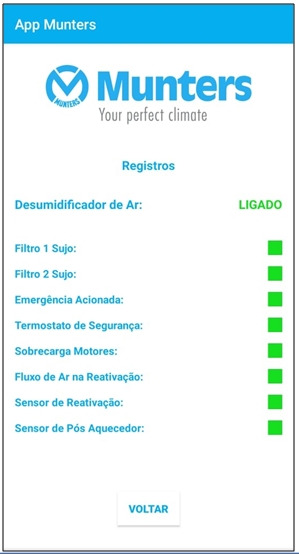} & \includegraphics[width=3.8cm]{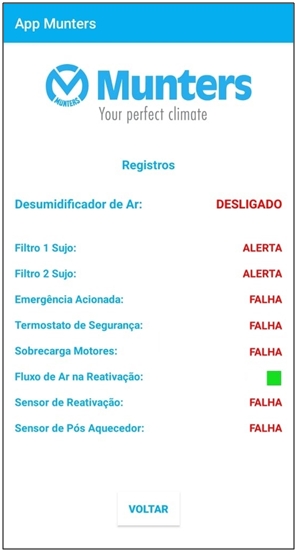} & \includegraphics[width=3.8cm]{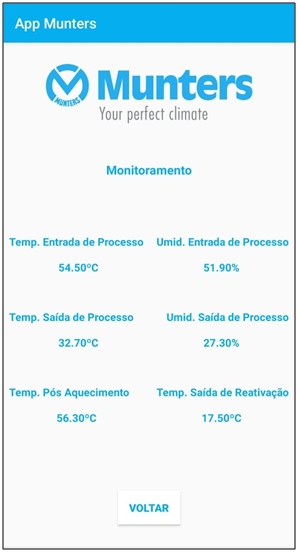} \\
		(a) & (b) & (c) \\
	\end{tabular}
	\caption{Application Mobile View: (a) ideal condition of dehumidifier, (b) conditions with fails and (c) measures parameters of dehumidifier.}
	\label{fig:llcd}
\end{figure*}

In Fig. 4a was simulated the initial test with the equipment in perfect working order, in which the equipment does not show any failure or active alarm.
When the machine detects a fault, automatically the equipment goes into shutdown mode, in Fig. 4b were simulated equipment failures in which activating the keys was caused the triggered emergency errors, safety thermostat, overhead motor and increasing the resistance of pot, generate the error sensor reactivation and post heater.
The reactivation in the air flow is operating for the reason that it only comes into fault with the connected equipment is failure refers to the differential pressure switch, equal to the filters, but it a failure because it is considered is located after the resistance and Therefore it is very important for the safety of equipment. 
Another function performed by the application is monitoring in real time Fig. 4c shows the values received from the PLC through temperature and humidity sensors, which are simulated by means of resistance change, which was used potentiometers.

The Fig. 4c shows the values in real time, as the process input is 54.50$^{\circ}$C temperature and after passing through the dehumidifier reaches the setting cylinder with a humidity of 51.90\%, the same applies to the process output the reactivation output reading sensor directly affects the process input moisture, as its basis that the PLC using as a reference for PID control. This information is important for the user to know how is the operation of the equipment and make the necessary changes according to the needs of the process.

\section{Conclusion}

In this paper was proposed an Embedded System for monitoring the conditions of industrial dehumidifiers.
The working and test of system was made at a bench created with the original materials and connections of the proposed system. 
Based in results, we analyses and validate the operation conditions of equipments, programing softwares and communication data.

As it has not been tested the embedded system at a real industrial environment, it is suggested to apply its on a real scenario on further works.
In addition to suggesting to create a iOS version of mobile application.

\end{document}